\documentclass[runningheads]{llncs}
\usepackage{amsmath,graphicx}
\usepackage{amssymb}
\usepackage[misc,geometry]{ifsym}

\begin{document}
\title{Boosting StarGANs for Voice Conversion with Contrastive Discriminator}
%
\author{Shijing Si\inst{1,2} \and Jianzong Wang\inst{1}\Letter\thanks{Corresponding author: Jianzong Wang, jzwang@188.com} \and Xulong Zhang\inst{1} \and Xiaoyang Qu\inst{1} \and Ning Cheng\inst{1} \and Jing Xiao\inst{1}}
\institute{Ping An Technology (Shenzhen) Co., Ltd., China \\
\and School of Economics and Finance, Shanghai International Studies University, China}
%
%
%

%
\maketitle
\begin{abstract}
Nonparallel multi-domain voice conversion methods such as the StarGAN-VCs have been widely applied in many scenarios. However, the training of these models usually poses a challenge due to their complicated adversarial network architectures. To address this, in this work we leverage the state-of-the-art contrastive learning techniques and incorporate an efficient Siamese network structure into the StarGAN discriminator. Our method is called SimSiam-StarGAN-VC and it boosts the training stability and effectively prevents the discriminator overfitting issue in the training process. We conduct experiments on the Voice Conversion Challenge (VCC 2018) dataset, plus a user study to validate the performance of our framework.
Our experimental results show that SimSiam-StarGAN-VC significantly outperforms existing StarGAN-VC methods in terms of both the objective and subjective metrics.
\end{abstract}

\section{Introduction}
\label{sec:intro}
Voice conversion (VC) is a speech processing task that converts an utterance from one speaker to that of another \cite{stylianou1998continuous,nercessian2020zero,zhou2021seen,zhang2021transfer}. 
VC can be useful to various scenarios and tasks such as speaker-identity modification
for text-to-speech (TTS) systems \cite{kim2020emotional}, speaking assistance \cite{urabe2020electrolarynx}, and speech enhancement \cite{chen2020enhancing}.

Voice contains significant information of the speaker \cite{si2021speech2video},
so increasingly complicated models are employed to capture the feature of voice.
Statistical methods based on Gaussian mixture models (GMMs) \cite{toda2007voice,helander2010voice} have been quite successful in VC task.
Recently, deep neural networks (DNNs), including
feed-forward deep NNs \cite{saito2017voice}, recurrent NNs \cite{sun2015voice}, and
generative adversarial nets (GANs) \cite{kaneko2018cyclegan}, have also achieved promising results on VC task. Most of these conventional VC methods require accurately aligned parallel source and target speech data.
However, in many scenarios, it may be impossible to access parallel utterances. Even if we could collect such data, we
typically need to utilize time alignment procedures, which
becomes relatively difficult when there is a large 
acoustic gap between the source and target speech \cite{hsu2016voice,kaneko2018cyclegan}. These challenges motivate how to train high-quality VC models with non-parallel data.

Many research study non-parallel VC methods, because they
require no parallel utterances, transcriptions, or time alignment procedures. Currently, two
representative methods of this type are CycleGAN-VCs \cite{kaneko2019cyclegan,kaneko2020cyclegan} and StarGAN-VCs \cite{kameoka2018stargan}. These methods are first developed by the computer vision (CV) community for style transfer of figures \cite{zhu2017unpaired,choi2018stargan}.
The main difference between CycleGAN-VCs and StarGAN-VCs lies in the multi-domain cases. CycleGAN-VCs are specialized to two domain cases, while StarGAN-VCs can handle multi-domains by taking account of the latent code for each domain \cite{kameoka2018stargan}. Other researchers also investigate how to perform voice coversion in few-shot cases, such as, \cite{tang2021tgavc} and \cite{tang2022avqvc}.
However, the training of GAN-like models is a challenge due to their non-convex nature. Therefore, the training stability of StarGAN-VCs is poor and can consume a significantly large amount of time.

In this paper, we focus on how to boost the training stability of StarGAN-VCs that utilize a StarGAN architecture to perform VC tasks.
Due to the non-convex/stationary nature of the mini-max game, however, training StarGANs in practice
is often very unstable and extremely sensitive to many hyperparameters \cite{salimans2016improved,choi2020stargan}. Data augmentation techniques have recently proven beneficial to stabilizing GAN-like adversarial models \cite{Zhang2020Consistency}.
Researchers also have applied contrastive learning methods to the basic GAN as an
auxiliary task upon the GAN loss \cite{lee2021infomax,jeong2021training}. From the literature, contrastive methods can strengthen the discriminator of GAN models, thereby improving the capability of the entire GAN model. However, little attempt has been done to the complicated StarGAN models, let alone for the VC tasks.
In this paper, we leverage the efficient simple Siamese (SimSiam) representation learning \cite{chen2021exploring}, one kind of contrastive learning, to train the discriminator of the StarGAN-VC model, and our method is called SimSiam-StarGAN-VC. We evaluated the performance of the proposed SimSiam-StarGAN-VC on the commonly used multi-speaker VC dataset
Conversion Challenge 2018 (VCC 2018) \cite{lorenzo2018voice}. We observe that SimSiam-StarGAN-VC presents better stability
of the training process and better naturalness of converted voices, compared with the original StarGAN-VC2.

Our contributions are summarized as follows:
\begin{itemize}
    \item We propose a SimSiam-StarGAN-VC method, which incorporates a Siamese network into StarGAN-VCs and stabilizes the training of StarGAN-VCs.
    \item We empirically investigate the performance of SimSiam-StarGAN-VC and show its superiority over StarGAN-VCs in terms of both subjective and objective metrics.
\end{itemize}

\section{Background}
\label{sec:back}
Prior to the introduction of our SimSiam-StarGAN-VC, we elaborate the StarGAN-VC2 and SimSiam methods in this section.

\subsection{StarGAN-VC2 Method}

Inspired by the success of StarGAN in the computer vision community, \cite{kameoka2018stargan} proposed to leverage its power to train a single generator $G$ that converts voices among multiple speakers or domains. 
For each speaker, StarGAN-VC posits a domain code (e.g., a speaker identifier or embedding).
The generator $G$ of StarGAN-VC takes a real acoustic feature map $\boldsymbol{x}$ and the target domain code $c'$ as input and produces a feature map $\boldsymbol{x}'$ of the target speaker
domain $c'$. The mathematical notations are presented in Table \ref{tab:notation}. Specifically,
We denote $\boldsymbol{x}$
as a 2-dimensional acoustic feature map (like MFCC). We use
$c \in \{1,\ldots, N\}$ to denote the domain code of a speaker, where the number of domains or speakers is $N$.

To further enhance the conversion performance of StarGAN-VC,
StarGAN-VC2 \cite{kaneko2019stargan} introduces the source-and-target conditional adversarial loss to replace the classification loss and target conditional adversarial loss in StarGAN-VC.
Both the generator and discriminator in StarGAN-VC2 take the source $(c)$ and target $(c')$ codes as input, i.e., $G(\boldsymbol{x}, c, c')\rightarrow{\boldsymbol{x}'}$. The training objectives of StarGAN-VC2 is the source-and-target adversarial loss which is shown as follows.

\noindent\textbf{Source-and-target adversarial loss:} the most significant contribution of StarGAN-VC2
\begin{equation}\label{eq:st_adv}
\begin{aligned}
\mathcal{L}_{st-adv} =&~\mathbb{E}_{(\boldsymbol{x}, c) \sim P(\boldsymbol{x}, c), c^{\prime} \sim P\left(c^{\prime}\right)}\left[\log D\left(\boldsymbol{x}, c^{\prime}, c\right)\right]+ \\
&~\mathbb{E}_{(\boldsymbol{x}, c) \sim P(\boldsymbol{x}, c), c^{\prime} \sim P\left(c^{\prime}\right)}\left[\log(1 - D\left(G\left(\boldsymbol{x}, c, c^{\prime}\right), c, c^{\prime}\right))\right],
\end{aligned}
\end{equation}
lies in that both the generator $G$ and discriminator $D$
takes the acoustic feature map ($\boldsymbol{x}$), source domain ($c$) and target domain codes ($c'$) as input. The striking difference between StarGAN-VC2 and StarGAN-VC is
that StarGAN-VC ignores the source code $c$.
$D(\boldsymbol{x}, c', c)$ outputs the probability that an acoustic feature $\boldsymbol{x}$ is real from the target domain $c$, and its range is from 0 to 1.
Similar to other GAN models, it is a min-max game:
maximizing the loss in Eq. \eqref{eq:st_adv} with respect to $D$ leads to a powerful fake voice detector, but minimizing the loss with regards to $G$ will train a generator to mimic the true acoustic features. 

 \cite{kaneko2019stargan} also explored to deploying a conditional instance normalization (CIN) module \cite{dumoulin2017learned} inside the network architecture, which proceeds as in Eq. \ref{eq:cin}. 
\begin{equation}\label{eq:cin}
\operatorname{CIN}\left(\boldsymbol{f}; c^{\prime}\right)=\gamma_{c^{\prime}}\left(\frac{\boldsymbol{f}-\mu(\boldsymbol{f})}{\sigma(\boldsymbol{f})}\right)+\beta_{c^{\prime}},
\end{equation}
In Eq. \ref{eq:cin}, $\boldsymbol{f}$ represents a feature map of input audio, $\mu(\boldsymbol{f})$ and $\sigma(\boldsymbol{f})$ are the average and standard deviation of $\boldsymbol{f}$ that are computed for each training sample. $\gamma_{c^{\prime}}$ and $\beta_{c^{\prime}}$ are speaker(domain)-specific scale and bias parameters for the speaker (i.e., domain) $c$.
In the training process, we train and learn these speaker-specific parameters with other network parameters/weights. For the source and target generator loss in Eq. \eqref{eq:st_adv}, these domain specific parameters $\gamma$ and $\beta$ are dependent on both the source ($c$) and target speakers ($c'$), i.e., $\gamma_{c^{\prime}}$ and $\beta_{c^{\prime}}$ are replaced by $\gamma_{c,c^{\prime}}$ and $\beta_{c,c^{\prime}}$, respectively.



\subsection{Simple Siamese Representation Learning}

SimSiam is one kind of contrastive learning \cite{chen2021exploring}, which requires none of the following: 1.) negative sample pairs, 2.) large batches, and 3.) momentum encoders. 
It utilizes two random data augmentations of each audio as input, and extracts features via the same encoder network $f$ and a multi-layer perceptron (MLP) projection header $h$. More specifically,
augmented speeches $\boldsymbol{x}_1$ and
$\boldsymbol{x}_2$ come from $\boldsymbol{x}$, with their high-level features $z_{i}=f(\boldsymbol{x}_{i})$ and
$p_{i}=h(f(\boldsymbol{x}_{i}))$.
The SimSiam loss for each real speech is
\begin{equation}\label{eq:simsiam.loss}
\begin{split}
        \mathcal{L}_{SimSiam}(\boldsymbol{x}_{1}, \boldsymbol{x}_{2}, f) = &\frac{1}{2}\mathcal{D}(p_1, \text{\fontfamily{qcr}\selectfont stopgrad}(z_2)) \\
        &+ \frac{1}{2}\mathcal{D}(p_2, \text{\fontfamily{qcr}\selectfont stopgrad}(z_1)),
\end{split}
\end{equation}
where $$\mathcal{D}(p_1, z_2)=-\frac{p_1}{||p_1||_2}\cdot\frac{z_2}{||z_2||_2}$$ with $||\cdot||_2$ the $\ell_2$-norm and $\text{\fontfamily{qcr}\selectfont stopgrad}$ represents the stop-gradient operation.



\begin{figure}[!t]
	\begin{center}
		\includegraphics[width=.6\textwidth]{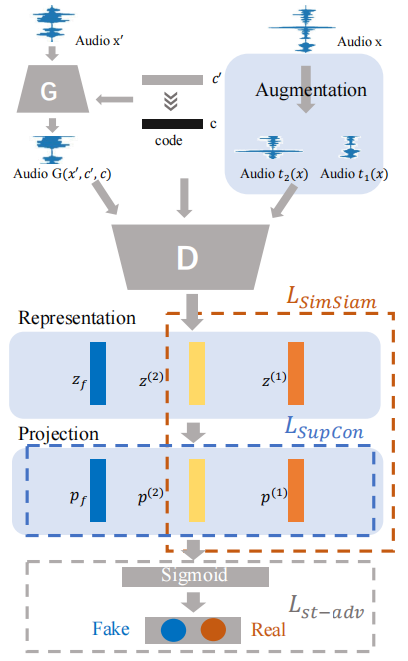}
		\caption{The overall architecture of SimSiam-StarGAN-VC. The source and target speaker (domain) codes are $c$ and $c'$, respectively.}
		\label{fig:star.gan.vc}
	\end{center}
\end{figure}%

\section{Methodology}
\label{sec:method}

In this section, we illustrate how our SimSiam-StarGAN-VC works. We utilize the same network architecture as StarGAN-VC2 in \cite{kaneko2019stargan}, but our framework is compatible to many existing GAN based VC architectures.
The overall architecture is shown in Fig. \ref{fig:star.gan.vc}, where the $G$ and $D$ are the generator and discriminator, respectively. The source and target domain codes are $c$ and $c'$, respectively, and they are embedded to latent vectors before being fed into the generator and discriminator.
For clarity, we list the mathematical notations in Table \ref{tab:notation}.

\begin{table}[h]
    \centering
    \caption{List of mathematical notations\label{tab:notation}}
    \begin{tabular}{l| l}
    \hline
        Notation & Meaning \\
         \hline
        $\boldsymbol{x}$ & MFCC features of a speech  \\
       $D_{e}$  & Encoder part of the discriminator $D$\\
       $d_{e}$  & The number of features in the output of $D_{e}$\\
        $t_1$, $t_2$ & Two kinds of data augmentation (DA)  \\
       $t_{1}(\boldsymbol{x})$ & MFCC features of a speech after DA $t_{1}$ \\
       $t_{2}(\boldsymbol{x})$ & MFCC features of a speech after DA $t_{2}$ \\ 
       $\boldsymbol{z}_{i}^{(1)}$ & \emph{i.e.}, $ D_{e}(t_{1}(\boldsymbol{x}_{i}))$, hidden features \\
       & of the $i$-th speech after DA $t_1$\\
       $\boldsymbol{z}_{i}^{(2)}$ & Same as $\boldsymbol{z}_{i}^{(1)}$, but for DA $t_2$\\
      $\boldsymbol{p}_{i}^{(1)}$ & Projected features of the i-th sample, \emph{e.g.},\\ 
      & the output of feeding $\boldsymbol{z}_{i}^{(1)}$ to a linear layer \\
      $\boldsymbol{p}_{i}^{(2)}$ & Same as $\boldsymbol{p}_{i}^{(1)}$, but for DA $t_2$ \\
      $B$ & The number of real samples in a training batch\\
      $\boldsymbol{p}^{(1)}$ & The set $\{\boldsymbol{p}^{(1)}_{i}, i=1, \ldots, B\}$ of projected \\
      &features of real samples after DA $t_1$  \\
      $\boldsymbol{p}^{(2)}$ & The set $\{\boldsymbol{p}^{(2)}_{i}, i=1, \ldots, B\}$ with DA $t2$ \\
      $P_{i+}^{(2)}$ & Subset of $\boldsymbol{p}^{(2)}$ containing samples of the same \\  
      & label (True or Fake) as the $i$-th sample \\
      $\boldsymbol{p}_{f,i}$ & The projected features of the $i$-th fake \\
      & (generated by $G$) audio sample \\
      $\boldsymbol{p}_{f,-i}$ & The set of projected features for all generated \\
      & samples except the $i$-th sample 
         \\\hline
    \end{tabular}
\end{table}

\subsection{Contrastive learning for real samples}

In this part, we describe how to train the discriminator $D$ with contrastive learning.
We denote the encoder part of the discriminator $D$ 
as $D_{e}$, and it can extract high-level features (a real vector) from an input speech, i.e., $D_{e}: (\boldsymbol{x})\rightarrow{\mathbb{R}^{d_{e}}}$.

 Overall, the
encoder network $D_{e}$ of SimSiam-StarGAN-VC is trained by minimizing two different contrastive losses: (a) the
SimSiam loss in Eq. \ref{eq:simsiam.loss} on the real speech samples, and (b) the supervised contrastive loss \cite{khosla2020supervised} on fake speech samples. Fig. \ref{fig:star.gan.vc} displays the loss functions used in our SimSiam-StarGAN-VC.
We elaborate these two contrastive losses in detail.

\noindent\textbf{Contrastive learning with real speech samples}
Here, we attempt to simply follow the SimSiam training scheme for each real sample $\boldsymbol{x}$, the loss function is
\begin{equation}\label{eq:simsiam}
    L_{Sim}(\boldsymbol{x}, D_{e}) = L_{SimSiam}(t_{1}(\boldsymbol{x}), t_{2}(\boldsymbol{x}), D_{e}),
\end{equation}
where $t_1$ and $t_2$ are augmentation methods for audio data
\cite{park2019specaugment}.

\subsection{Supervised contrastive learning for fake speech samples} In order for the encoder of $D_{e}$ to keep
necessary information to discriminate real and fake speech samples, we consider an auxiliary loss $L_{con}$. Specifically, we employ the supervised contrastive loss \cite{khosla2020supervised} over fake (generated) speech samples. This loss is an extended version of contrastive loss to support supervised learning by allowing
more than one sample to be positive, so that samples of the same label can be attracted to each other in
the embedding space. On a mini-batch, we treat the real samples and their augmented versions as positive, and the generated fake speech samples as negative. For a mini-batch of real samples, we denote $\boldsymbol{p}^{(1)}$ and $\boldsymbol{p}^{(2)}$ as the projected representations (after a MLP) of two kinds of data augmentation. $\boldsymbol{p}_{f}$ is the set of projected representations for a batch of converted (generated by $G$) fake audio samples. 
Formally, for each $\boldsymbol{p}_{i}^{(1)}$, let $P_{i+}^{(2)}$ be a subset of $\boldsymbol{p}^{(2)}$ that represent the positive pairs for $\boldsymbol{p}_{i}^{(1)}$. Then the supervised contrastive loss is defined by:
\begin{equation}\label{eq:super.con}
\begin{split}
        &L_{SupCon}(\boldsymbol{p}_{i}^{(1)}, \boldsymbol{p}^{(2)}, P_{i+}^{(2)})=\\
        &-\frac{1}{|P_{i+}^{(2)}|}\sum_{\boldsymbol{p}_{i+}^{(2)}\in{P}_{i+}^{(2)}}\log\frac{\exp(s(\boldsymbol{p}_{i}^{(1)},\boldsymbol{p}_{i+}^{(2)}))}{\sum_{j}\exp(s(\boldsymbol{p}_{i}^{(1)},\boldsymbol{p}_{j}^{(2)}))},
\end{split}
\end{equation}
where $s(\cdot, \cdot)$ is the inner product used in SimCLR \cite{chen2020simple}.

Using this notation, we define the loss for fake samples as follows:
\begin{equation}
L_{Con} =\frac{1}{B}\sum_{i=1}^{B}L_{SupCon}(\boldsymbol{p}_{f,i}, [\boldsymbol{p}_{f,-i}; \boldsymbol{p}^{(1)};\boldsymbol{p}^{(2)}], [\boldsymbol{p}_{f,-i}]),
\end{equation}
where $B$ is the batch size and       $[\boldsymbol{p}_{f,-i}; \boldsymbol{p}^{(1)};\boldsymbol{p}^{(2)}]$ is the union of three sets of projected features.

The loss function of SimSiam-StarGAN-VC for the generator is $L_{st-adv}$ in Eq. \eqref{eq:st_adv}, and for discriminator the loss is defined as follows:
\begin{equation}\label{eq:loss.simsiam}
    L_{D} = -L_{st-adv}+\lambda_1\cdot {L}_{Sim} +\lambda_2\cdot L_{Con},
\end{equation}
where $\lambda_1$ and $\lambda_2$ are the strength parameters for the SimSiam and supervised contrastive loss.



\section{Experiments}
\label{sec:exp}


\subsection{Experimental setup}
\label{exp.setup}
\noindent\textbf{Dataset:}
We utilized data from the most popular VCC 2018 dataset \cite{lorenzo2018voice}
in a similar manner to the experiments in StarGAN-VC2 \cite{kaneko2019stargan}. We describe how we conduct the experiments briefly.
To perform both inter-gender and
intra-gender VC, we randomly selected two male and two female speakers from VCC 2018, denoted as SF1, SF2, SM1,
and SM2, short for ``Speaker of Female/Male 1 or 2''.
Therefore, we have $N = 4$ as the number of domains (speakers). To ensure the non-parallel setting, there is no overlapping content between the training and evaluation datasets. 
For a thorough comparison, we conduct all $4 \times 3 = 12$ combinations intra-gender and inter-gender conversions.
Each speaker has approximately 80 utterances for model training and 30 for model evaluation.


\noindent\textbf{Implementation details:}
For StarGAN-VC and StarGAN-VC2, we employ the same
network architecture as
shown in the Figure 3 of \cite{kaneko2019stargan}. 
For the data augmentation methods in the SimSiam-StarGAN-VC, we utilized time masking and frequency masking as $t_1$ and $t_2$, respectively.
The upper limit of training epochs is set to be $1\times 10^5$, early stopping is deployed, and the learning
rate parameters for $G$ and $D$ is tuned by carefully by closely monitoring the loss of discriminators and generators. 



\subsection{Objective evaluation}

As common in the literature, an objective evaluation is done to verify the benefits of our SimSiam-StarGAN-VC over other existing StarGAN-VCs. Similar to \cite{kaneko2019stargan}, we also utilized the Mel-cepstral distortion (MCD) and the modulation spectra distance (MSD).
Essentially, these two metrics
measure the overall and local structural
differences between the target and converted Mel-cepstral coefficients (MCEPs). For both MCD and MSD metrics,
smaller values indicate better voice conversion performance.

\begin{table}[!h]
\caption{Comparison of MCD and MSD among three different models.
\label{tab:obj}}
\begin{center}
\begin{tabular}{c|ccc}
\hline
Method &   & MCD [dB]~~  & MSD [dB]  \\[2pt]\hline
StarGAN-VC & & $7.11\pm.10$~~ &$2.41\pm.13$\\[2pt] 
StarGAN-VC2 & &$6.90\pm.07$~~  &$1.89\pm.03$\\[2pt]\hline 
SimSiam-StarGAN-VC & &$\textbf{6.35}\pm\textbf{.12}$~~  & $\textbf{1.48}\pm\textbf{.10}$\\[2pt]
\hline
\end{tabular}
\end{center}
\end{table}
Table \ref{tab:obj} displays the performance
of 3 different VC approaches in terms of two objective metrics (MCD and MSD).
To show the statistical significance, we have computed the mean scores by taking average over models trained with five different initializations and reported the standard deviations in this table. From Table \ref{tab:obj}, our SimSiam-StarGAN-VC (MCD: 6.35, MSD: 1.48) significantly outperforms both StarGAN-VC (MCD: 7.11, MSD: 2.41)
and StarGAN-VC2 (MCD: 6.90, MSD: 1.89) in terms of both metrics. This indicates that contrastive losses ($L_{Sim}$ and $L_{Con}$) are useful for improving the feature extraction capability of the discriminator, which further boosts the quality of converted speeches.


\subsection{Subjective evaluation}

To analyze the effectiveness of SimSiam-StarGAN-VC, we conducted listening tests to cpmpare it with StarGAN-VC2. We collected 36 generated (converted) sentences (12 source-target combinations $\times 3$ sentences, where the first one is the real target utterance and the other two are generated by SimSiam-StarGAN and StarGAN-VC2). Eight well-educated Chinese native speakers participated in the tests as audiences. We conducted a mean opinion score (MOS) test to evaluate the naturalness of generated speeches, from 5 (for excellent quality) to 1 (for poor quality). In these tests we presented the target speech as a reference for audiences (the average MOS for target speeches is about 4.5), so that the audiences can evaluate the generated speeches properly.

We also implemented an XAB test to evaluate speaker similarity by randomly selecting 30 sentences from the evaluation set.
Here we denote ``X" the target speech,
and ``A" and ``B" were converted utterances from StarGAN-VC2
and SimSiam-StarGAN-VC, respectively. When presenting each set of speeches, we display ``X'' first, then ``A'' and ``B'' randomly.
After the audiences heard one set of speeches, we asked them to choose which speech (``A'' or ``B'') is closer to the target (``X''), or to be ``Fair''.

\begin{figure}[!h]
	\begin{center}
		\includegraphics[width=\linewidth]{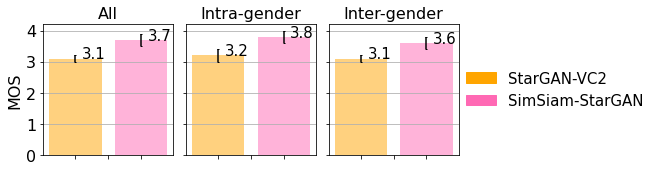}
		\caption{The average MOS values of all, intra-gender and cross-gender conversion of StarGAN-VC2 (orange bars) and SimSiam-StarGAN-VC (pink bars)}
		\label{fig:mos}
	\end{center}
\end{figure}

\begin{figure}[!h]
	\begin{center}
		\includegraphics[width=1.2\linewidth]{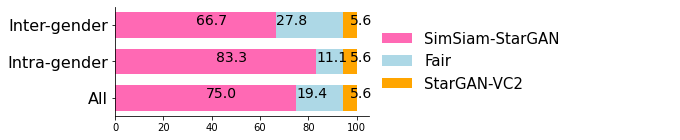}
		\caption{The preference of all, intra-gender and cross-gender conversion of StarGAN-VC2 (orange) and SimSiam-StarGAN-VC (pink)}
		\label{fig:pref}
	\end{center}
\end{figure}

Fig. \ref{fig:mos} and Fig. \ref{fig:pref} display the main findings of naturalness and the preference scores of StarGAN-VC2 and SimSiam-StarGAN-VC, respectively.
In Fig. \ref{fig:mos}, the pink and orange bars represent the MOS of SimSiam-StarGAN-VC and StarGAN-VC2, respectively.
These results empirically demonstrate that SimSiam-StarGAN-VC (overall MOS: 3.7) outperforms StarGAN-VC2 (overall MOS: 3.1) on naturalness for every category. In Fig. \ref{fig:pref}, the pink, light blue and orange colors the preference scores for SimSiam-StarGAN-VC, fair, and StarGAN-VC2, respectively. The SimSiam-StarGAN-VC (overall preference: 75.0\%) outperforms StarGAN-VC2 (overall: 5.6\%) significantly on speaker similarity.
We also highlight that our SimSiam-StarGAN-VC takes only $100$ training epochs to converge, which is shown in Fig. \ref{fig:train.vc}. However, StarGAN-VC2 still oscillates significantly after 400 epochs. This demonstrates the effectiveness of contrastive training of the discriminator.




 \subsection{Training Stability}
\begin{figure}[!h]
	\centering	\includegraphics[width=.8\textwidth]{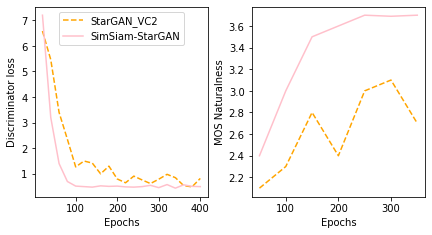}
		\caption{The discriminator loss and MOS values along training epochs. Left panel shows the discriminator loss of StarGAN-VC2 (orange dashed line) and SimSiam-StarGAN (pink solid line) versus the training epochs; and the right panel displays the MOS for naturalness of the two approaches.}
		\label{fig:train.vc}
\end{figure}

To show the training stability of SimSiam-StarGAN-VC, we have produced a figure of discriminator loss and mean opinion score (MOS) for speech naturalness along training epochs, which is presented in Fig. \ref{fig:train.vc}.

Fig. \ref{fig:train.vc} displays the discriminator loss traces and MOS for naturalness of StarGAN-VC2 (orange dashed lines) and SimSiam-StarGAN (pink solid lines). The discriminator loss of StarGAN-VC2 oscillates over training epochs, while the discriminator loss of SimSiam-StarGAN converges steadily. The right panel illustrates the MOS for naturalness of generated speech by these two methods. Similar to the left panel, the speech converted by SimSiam-StarGAN exhibits increasing MOS over epochs.

\subsection{Ablation Study on contrastive losses} 
We conducted comparative studies on the sensitivity of the hyperparametes $\lambda_1$ and $\lambda_2$ for the SimSiam-StarGAN-VC.
Table \ref{tab:ablation} exhibits the MCD scores over different combinations of $\lambda_1$ and $\lambda_2$, and $\lambda_1=\lambda_2=0.01$ is the best choice for the VCC 2018 dataset.
 We have recorded the MOS values for the ablation study in the experiments. The extended table in shown in Table \ref{tab:ablation}. When $\lambda_1 = \lambda_2 =0.01$, the SimSiam-StarGAN-VC performs best in terms of both the MCD and MOS for naturalness. We will include this table in the final version paper.

\begin{table}[h]
\caption{Ablation study of hyper-parameters $\lambda_1$ and $\lambda_2$.\label{tab:ablation}}
\centering
\begin{tabular}{l|llllll}
\hline
$\lambda_1$     & 0.0  & 0.01 & 0.01 & 0.02 & 0.05 & 0.1  \\
$\lambda_2$     & 0.01 & 0.0  & 0.01 & 0.05 & 0.02 & 0.1  \\ \hline
MCD{[}dB{]} & 7.23 & 6.56 & \textbf{6.35} & 6.48 & 6.55 & 6.95 \\
MOS & 3.05 & 3.56 & \textbf{3.70} & 3.68 & 3.65 & 3.45 \\
\hline
\end{tabular}
\end{table}

\section{Conclusion}
\label{sec:conclusion}

To advance the research on multi-domain non-parallel voice conversion, we have incorporated the contrastive learning methods in StarGAN-VC during the training stage. We leveraged the SimSiam and supervised contrastive loss to enhance the capability of the encoder of the discriminator.
The empirical studies on non-parallel multi-speaker VC demonstrate
the effectiveness of our SimSiam-StarGAN-VC. 
Therefore, contrastive learning methods can boost the performance of StarGANs on the VC task by improving the convergence and stability of the complicated StarGAN training.
Contrastive learning has shown good promise in the computer vision community. It is reasonable to believe that it will advance the speech processing area in many aspects. In the next step, we may attempt to employ the variational information bottleneck \cite{si2021variational} with contrastive learning to disentangle the speaker identity information from the input speech, which may improve the controllability of VC models.

\section{Acknowledgment}
This paper is supported by the Key Research and Development Program of Guangdong Province under grant No.2021B0101400003. Corresponding author is Jianzong Wang from Ping An Technology (Shenzhen) Co., Ltd (jzwang@188.com).

\bibliographystyle{splncs04}
\bibliography{refs}

\end{document}